\documentclass[epjCONF,final,columns]{svjour}
\usepackage[dvipdfmx]{graphics}
\usepackage[varg]{txfonts} % Times fonts
\usepackage[latin1]{inputenc}
\session-title{}

\begin{document}
\title{Fingering induced by a solid sphere impact to viscous fluid}
\author{Hiroaki Katsuragi\thanks{\email{katsurag@eps.nagoya-u.ac.jp}} }
\institute{Department of Earth and Environmental Sciences, Nagoya University, Furocho, Chikusa, Nagoya 464-8601, Japan}
\abstract{
The number of splashed fingers generated by a solid projectile's impact onto a viscous liquid layer is experimentally studied. A steel sphere is dropped onto a viscous liquid pool. Then, a fingering instability occurs around the crater's rim, depending on the experimental conditions such as projectile's inertia and the viscosity of the target liquid. When the impact inertia is not sufficient, any fingering structure cannot be observed. Contrastively, if the impact inertia is too much, the random splashing is induced and the counting of fingers becomes difficult. The clear fingering instability is observable in between these two regimes. The number of fingers $N$ is counted by using high-speed video data. The scaling of $N$ is discussed on the basis of dimensionless numbers. By assuming Rayleigh-Taylor instability, scaling laws for $N$ can be derived using Reynolds number $Re$, Weber number $We$, and Froude number $Fr$. Particularly, the scaling $N=(\rho_r Fr)^{1/4}We^{1/2}/3^{3/4}$ is obtained for the gravity-dominant cratering regime, where $\rho_r$ is the density ratio between a projectile and a target. Although the experimental data considerably scatters, the scaling law is consistent with the global trend of the data behavior. Using one of the scaling laws, planetary nano crater's rim structure is also evaluated. 
} %end of abstract
\maketitle
\section{Introduction}
Since Worthington's pioneering work for the splashing of fluid impact~\cite{Worthington}, the impact splashing has been studied intensively by many physicists. Recent developments of the numerical computation and the high-speed imaging enable us to reveal the detailed dynamics of the fluid impact~\cite{Yarin,Clanet,ThoroddsenRev}. Particularly, a liquid droplet impact either onto a liquid pool or a hard floor has been studied well so far~\cite{Yarin}. A solid sphere impact to a non-Newtonian target fluid has been also investigated experimentally~\cite{Tabuteau,Ara}. Furthermore, some recent studies have concerned granular target impacts as well as fluid target impacts. Examples include a solid sphere's impact to a granular bed~\cite{Uehara,Walsh,Katsuragi2007,Katsuragi2013,Marston,Royer}, a liquid droplet impact to a granular bed~\cite{Katsuragi,Katsuragi2}, and a vortex ring impact to a granular bed~\cite{Sano}. 

In spite of these recent efforts for the soft matter impact, fingering instability by the impact between a solid sphere and a liquid pool has not been studied well. This is a little surprising because it is probably one of the simplest setups. So far, the study of fingering instability has been almost restricted within a droplet impact. However, natural fingering instability is not limited in the macroscopic droplet impact. For instance, microscopic impact cratering has been found in the space environment. Nakamura et al. have found wavy rim structures of tiny ($\simeq 100$ nm order) craters on the sample returned from the asteroid Itokawa~\cite{Nakamura}. This crater rim structure reminds us the fingering instability. The petal-like crater's rim structure has been also found in the droplet impact to a granular bed~\cite{Katsuragi,Katsuragi2}. From the high-speed video data taken in the experiments, we have confirmed that the petal-like structure is caused by the fingering instability of the impacting droplet. Petal-like ejecta blanket structure called Rampart crater can be also found in the Martian surface and a laboratory experiments of the hypervelocity impact to a muddy target~\cite{Rampart}. Such simple laboratory experiments might be useful to discuss the origin of widely variated natural craters shapes. Note that, however, surfaces of actual astronomical objects do not consist of liquid. They are rather covered with regolith. Moreover, Itokawa's returned samples, on which nano craters were found, are solid (not liquid). Thus the real craters morphologies might not directly link to the liquid impact. However, we believe that the fundamental study of the liquid impact could be a first step to understand the variety of complex craters morphologies.

To classify the physical origins of various fingering-like structures in soft matter impacts, fundamental investigations on the fingering instability with various projectiles and targets are necessary. For example, viscoelastic property of the target material might play a certain role in the very high-speed impact in which the melting of the target could occur. As a first step, we should begin with the simplest setup. Therefore, we carry out one of the simplest impact experiments---a steel sphere impact onto a viscous liquid---in order to approach a fundamental law of the fingering instability. Particularly, we analyze the experimental data by combining the models of liquid deformation and Rayleigh-Taylor instability. Using the derived model, we speculate the impact velocity to make astronomical nano craters accompanied with wavy rim structure. Then the limitations for this consideration are also discussed finally. 

\section{Experimental}
\begin{figure}
\begin{center}
\scalebox{1.15}[1.15]{\includegraphics{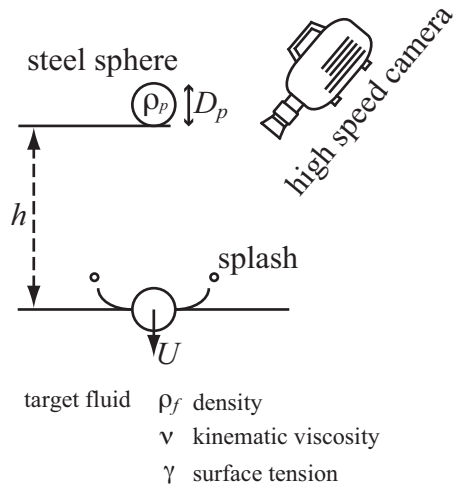}}
\end{center}
\caption{A schematic drawing of the experimental apparatus. A stainless steel sphere is dropped onto a viscous liquid pool by free fall. The splashing is taken by a high-speed camera.}
\label{fig:f1}
\end{figure}

We build a simple experimental apparatus as schematically shown in figure~\ref{fig:f1}. A steel sphere in diameter $D_p=3$, $6.35$, or $8$ (mm) is held by an electromagnet. Then it is dropped from a certain height $h$ to commence a free fall. The free-fall height $h$ ranges from $20$ to $605$ (mm) corresponding to the impact velocity of $U=0.2-3.6$ (m/s). This free-fall drop system is basically identical to that used in our previous experiment on the agar gel impact~\cite{Ara}. Density of the steel sphere is $\rho_p=8.0 \times 10^3$ (kg/m$^3$). We employ silicone oil (Shin-Etsu Chemical Co., Ltd.) or distilled water as target fluids. The range of kinematic viscosity $\nu$ and surface tension $\gamma$ of target fluids are $0.65 \leq \nu \leq 200$ ($\times 10^{-6}$ m$^2$/s$(=$cSt)) and $16 \leq \gamma \leq 72$ ($\times 10^{-3}$ N/m), respectively. The density of fluid ranges $0.8 \leq \rho_f \leq 1.0$ ($\times 10^3$ kg/m$^3$). Namely, main control parameters are the impact inertia and the target viscosity. Target surface tension and density are varied slightly. Actually, we have used the viscoelastic fluid target as well. However, we could not observe any fingering instability for viscoelastic targets, at least in the range of current experimental condition. Much larger impact inertia is probably required to induce the fingering instability in the viscoelastic impact. Thus we are going to focus only on the viscous target case, in this study. Impact and splashing are captured by a high-speed camera (Photron SA-5) at $5,000$ fps. All the experiments are performed under the atmospheric pressure environment ($\simeq 101$ kPa).

\section{Results and discussion}
\begin{figure}
\begin{center}
\scalebox{0.9}[0.9]{\includegraphics{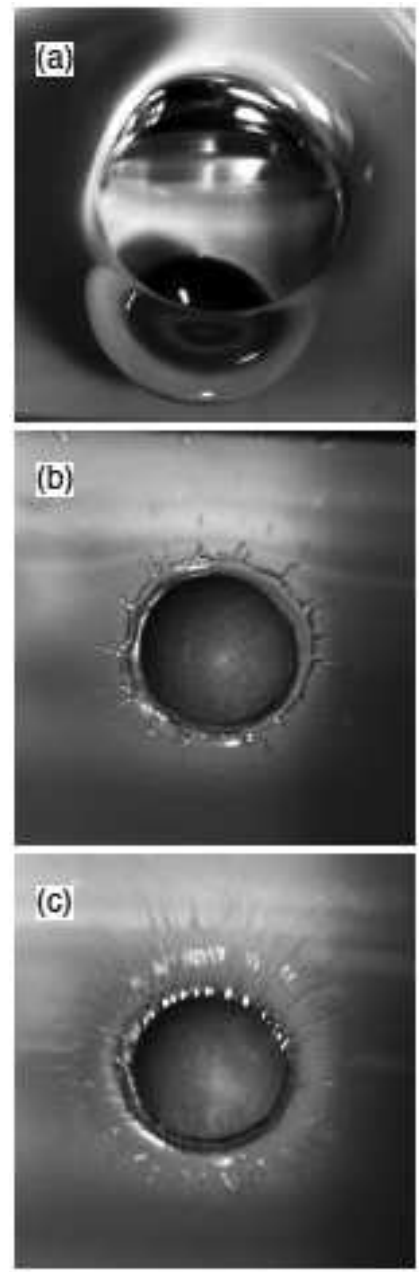}}
\end{center}
\caption{Typical snapshots of the impacts: (a) no-fingering case ($d=8$ mm, $\nu= 120$ cSt, and $h=240$ mm); (b) fingering case ($d=6.35$ mm, $\nu=0.89$ cSt, and $h=120$ mm); (c) random splashing ($d=6.35$ mm, $\nu =0.65$ cSt, and $h=605$ mm).}
\label{fig:f2}
\end{figure}

Typical snapshots of the experimental results are presented in figure~\ref{fig:f2}. When the target viscosity is large, any fingering instability cannot be observed. Such an example ($\nu=120$ cSt) is shown in figure~\ref{fig:f2}(a). To induce the fingering instability on a very viscous target, much larger impact inertia is necessary. On the other hand, we can confirm the clear fingering instability for an impact to a less viscous target (figure~\ref{fig:f2}(b); $\nu = 0.89$ cSt). When the impact inertia increases more, random splashing is induced as shown in figure~\ref{fig:f2}(c). In this regime, it is hard to count the number of fingers from the raw video data. Perhaps, this limit comes from the limitations of temporal and spatial resolutions of the images taken by the high-speed camera. Much higher resolution images might enable us to study this random regime. Since we are interested in the typical fingering structure, we restricted ourselves to the transient regime in which the clear fingering structure can be observed. 

%figure

Number of fingers $N$ is counted from the snapshots (like figure~\ref{fig:f2}(b)) by eye. Since the fingers move, split, and merge, it is not so easy to count $N$. Moreover, the fingering structure is finally relaxed relatively in a short time. To see the initial instability, we count all of possible fingers at the early stage of the impact. We would like to characterize $N$ by the scaling analysis using some relevant dimensionless numbers. 

%figure
\begin{figure*}
\begin{center}
\scalebox{1}[1]{\includegraphics{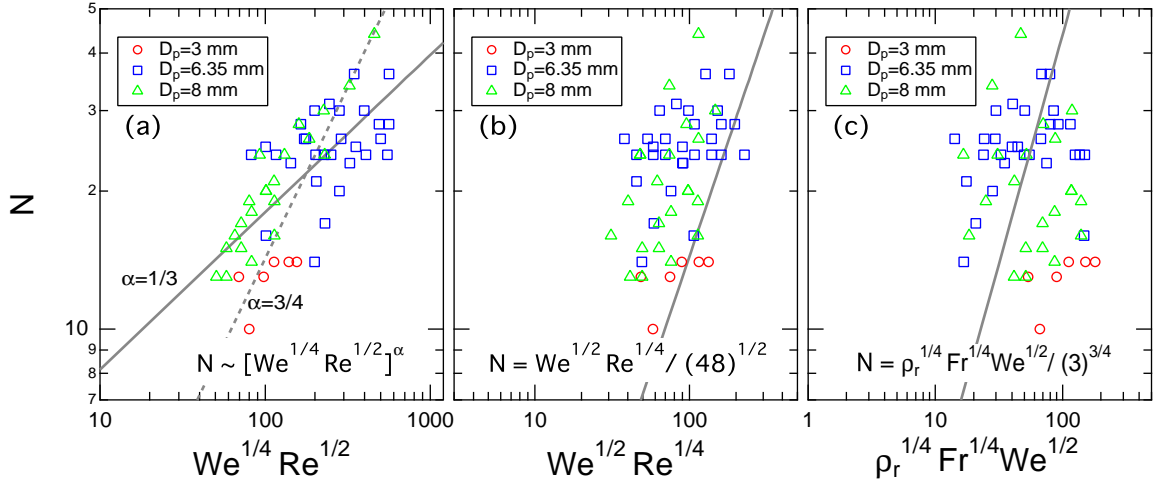}}
\end{center}
\caption{Scaling results of the counted fingers number $N$ by (a) impact Reynolds number $Re_I$, (b) droplet-deformation-included Rayleigh-Taylor instability, and (c) crater-deformation-based Rayleigh-Taylor instability. The solid and dashed lines in (a) indicate the scaling $N \sim {Re_I}^{1/3}$ and $N\sim {Re_I}^{3/4}$, respectively. The solid line in (b) represents $N= We^{1/2}Re^{1/4}/4\sqrt{3}$. The solid line in (c) corresponds to $N=(\rho_r Fr)^{1/4}We^{1/2}/3^{3/4}$. While these scaling laws capture the global trend of data behavior, the data dispersion is significantly large.}
\label{fig:f3}
\end{figure*}

\subsection{Impact Reynolds number}
First, the simple scaling analysis is tried by combining two possibly relevant dimensionless numbers: Weber number $We = \rho_f U^2 D_p/\gamma$ and Reynolds number $Re=UD_p/\nu$. Here we introduce the impact Reynolds number~\cite{Marmanis}. Impact Reynolds number has been derived by considering the representative length scale $L=[\nu (m_p/\gamma)^{1/2}]^{1/2}$, where $m_p$ is the sphere's mass. The characteristic timescale $t_{\gamma}=(m_p/\gamma)^{1/2}$ comes from the spring-like effect of the liquid surface tension. And $L$ indicates the length scale of momentum diffusion by $t_{\gamma}$ and $\nu$. Then, the impact Reynolds number $Re_I$ is defined as 
\begin{equation}
Re_I = \frac{UL}{\nu} = We^{1/4}Re^{1/2}.
\label{eq:Impact_Reynolds_numer}
\end{equation}
Marmanis and Thoroddsen have shown that the number of fingers by a droplet impacting to a hard floor can be scaled by $N\sim {Re_I}^{3/4}$~\cite{Marmanis}. In figure~\ref{fig:f3}(a), the counted result of $N$ as a function of $Re_I$ for the current experiment is displayed. Although the data considerably scatter due to the difficulty of fingers-counting, here we apply the data fitting to $Re_I$ scaling. Then, as shown in figure~\ref{fig:f3}(a), the data can be fitted by a scaling, 
\begin{equation}
N \sim {Re_I}^{\alpha}.
\label{eq:number_scaling}
\end{equation} 
The obtained scaling exponent $\alpha=1/3$ is different from the previously obtained droplet impact experiment, $\alpha=3/4$~\cite{Marmanis}. Solid and dashed lines in Fgi.~\ref{fig:f3}(a) correspond to $N\sim{Re_I}^{1/3}$ and $N\sim{Re_I}^{3/4}$, respectively. This difference might come from the geometry difference. While $\alpha=3/4$ was obtained for the impacting droplet case, $\alpha=1/3$ corresponds to the case that a hard projectile impact to a liquid target. Moreover, the obtained scaling means very weak $We$ and $Re$ dependences: $N \sim We^{1/12}Re^{1/6}$. Such small exponents are not so useful for the scaling analysis. It is hard to discuss the underlying physical mechanism on the basis of these small exponents. The data can be fitted only because the model of Eq.~(\ref{eq:number_scaling}) has two free fitting parameters. 

\subsection{Rayleigh-Taylor instability}
Next, Rayleigh-Taylor instability is considered as an alternative candidate to explain the fingering. Rayleigh-Taylor instability is induced at an interface of denser and lighter fluids when the denser one is  accelerated toward the lighter one. The capillary-based characteristic wave number $k$ for the Rayleigh-Taylor instability can be computed as~\cite{ChandrasekharFluid,Piriz2006,Bret2011}, 
\begin{equation}
k=\sqrt{\frac{(\rho_h - \rho_l)a}{3\gamma}},
\label{eq:RT_wavenumber}
\end{equation}
where $\rho_h$, $\rho_l$, and $a$ are the density of denser fluid, that of lighter fluid, and the applied acceleration, respectively. By assuming $\rho_f = \rho_h \gg \rho_l$ and $a \simeq U^2/D_p$, $k$ can be rewritten as $k=We^{1/2}/\sqrt{3}D_p$. If this instability is induced at the perimeter of a circle in diameter $D_c$, the number of fingers $N$ is written as,
\begin{equation}
N = \frac{D_c}{2\sqrt{3}D_p}We^{1/2}.
\label{eq:RT_N}
\end{equation}

By considering a liquid droplet impact onto a hard floor, the maximally deformed droplet diameter $D_c$ at viscosity-dominant regime can be approximated by~\cite{Mundo1995,Pasandideh-Fard1996,Bhola},
\begin{equation}
D_c \simeq \frac{1}{2} D_p Re^{1/4}.
\label{eq:Bhola_diameter}
\end{equation} 
Here, $D_p$ corresponds to the initial droplet diameter. Substituting Eq.~(\ref{eq:Bhola_diameter}) to Eq.~(\ref{eq:RT_N}), one can obtain a form of $N$ as~\cite{Bhola}, 
\begin{equation}
N= \frac{1}{4\sqrt{3}}We^{1/2}Re^{1/4}.
\label{eq:Bhola_N}
\end{equation}
It should be noticed that this scaling variable $We^{1/2}Re^{1/4}$ is different from $Re_I(= We^{1/4}Re^{1/2})$. The $We^{1/2}Re^{1/4}$ is sometimes called the splashing parameter. The scaling form (Eq.~(\ref{eq:Bhola_N})) is compared with the current experimental result in figure~\ref{fig:f3}(b). As can be seen, this scaling seems to correspond to the lower limit of the experimental data. Although the scaling in figure~\ref{fig:f3}(b) might look worse than that in figure~\ref{fig:f3}(a), note that the scaling of Eq.~(\ref{eq:Bhola_N}) does not include any fitting parameter. 

To improve the scaling more, here we consider the crater cavity model~\cite{Engel1966,Engel1967}. Let us assume that the projectile's kinetic energy $E_k$ is mainly consumed by the potential energy of the crater's cavity $E_{\rm cav}$. The cavity potential energy $E_{\rm cav}$ is calculated as $E_{\rm cav} = \pi \rho_f g D_c^4/64$~\cite{Engel1966}. By equating $E_{\rm cav}$ and the impacting kinetic energy $E_k=\pi \rho_p D_p^3 U^2/12$, the cavity diamater $D_{c}$ is obtained as,
\begin{equation}
D_{c} = 2\left( \frac{\rho_p D_p^3 U^2}{3\rho_f g} \right)^{1/4}=2D_p\left( \frac{\rho_p}{3\rho_f}Fr\right)^{1/4}.
\label{eq:cavity_radius}
\end{equation}
Substituting Eq.~(\ref{eq:cavity_radius}) to Eq.~(\ref{eq:RT_N}), a scaling form is obtained as, 
\begin{equation}
N= \left( 3^{-3} \rho_r Fr \right)^{1/4}We^{1/2},
\label{eq:Engel_N}
\end{equation}
where $Fr = U^2/gD_p$ is Froude number and $\rho_r={\rho_p}/{\rho_f}$ is the density ratio between a projectile and a target. This form looks similar to Eq.~(\ref{eq:Bhola_N}); $Re$ is replaced by $\rho_r Fr$ in terms of the scaling relation. While the dominant effect in Eq.~(\ref{eq:Bhola_N}) is viscosity, the gravity plays an essential role in Eq.~(\ref{eq:Engel_N}). This implies that the scaling of Eq.~(\ref{eq:Engel_N}) is valid for the gravity-dominant cratering regime. In figure~\ref{fig:f3}(c), the experimental data are compared with this scaling.  The abscissa of figure~\ref{fig:f3}(c) comes from the right-hand side of Eq.~(\ref{eq:Engel_N}). Whereas the data still widely scatter, the scaling captures the global trend without any fitting parameter. Perhaps, the small projectile case ($D_p=3$ mm) can be better explained by the viscosity-based scaling (Eq.~(\ref{eq:Bhola_N})). Other data seem to agree with the gravity-dominant scaling. This fingering relates to the surface tension effect and we did not directly measure the surface tension value (i.e. we use catalogue data). Therefore the data contain large uncertainty. That is probably the reason why it is hard to reduce the relatively large data dispersion in all panels of figure~\ref{fig:f3}. Moreover, the range of $We$ swept by this study is not wide enough. To make the scaling relation sure, more systematic experiments with widely ranging dimensionless numbers must be carried out. 

\subsection{Application to the fingering of nano craters found on the asteroid Itokawa}
In principle, we can estimate the impact conditions for various (gravity-dominant) craters with fingering instability, by using Eq.~(\ref{eq:Engel_N}). However, it is not easy to estimate the proper value of $Fr$ particularly for the astronomical impacts. Instead, the crater dimensions are directly accessible as far as the craters are observed. In such a situation, Eq.~(\ref{eq:RT_N}) is useful to evaluate the impact conditions. This scaling is based only on the Rayleigh-Taylor instability. It does not assume the dominant dynamics of the deformation: e.g. viscosity, gravity, or strength.

Here, let us estimate the impact velocity $U$ for the wavy rim structure of nano craters found on Itokawa's returned samples~\cite{Nakamura}. To this purpose, Eq.~(\ref{eq:RT_N}) is rewritten as, 
\begin{equation}
U=\frac{2N}{D_c} \sqrt{\frac{3D_p \gamma}{\rho_f}}. 
\label{eq:U_RT}
\end{equation}
In order to compute a specific value of $U$, here we use physical properties of SiO$_2$: $\rho_f =2000$ kg/m$^3$ and $\gamma=0.014$ N/m~\cite{Blum}. In addition, $N\simeq 7$ and $D_c\simeq 200$ nm can be observed from the photos of nano craters~\cite{Nakamura}. Assuming $D_p=D_c/2$, which is a reasonable assumption for fluid impacts, we finally obtain $U \simeq 100$ m/s. If we assume a molten glass state for the target, $\gamma$ value is about one order of magnitude greater than the currently assumed value. However, the resultant $U$ still remains in the order of $U = 10^2$ m/s. The assumption $D_p=D_c/2$ actually means $Fr\simeq 1$ if we use Eq.~(\ref{eq:cavity_radius}). However, it is quite difficult to satisfy this condition under the very weak gravity like Itokawa. Thus, the crater diameter $D_c$ must be determined rather by the target strength $Y$. By considering the energy balance between $E_{\rm cav,Y}=\pi D_c^3 T/12$ and $E_k$, $D_c$ is written as,
\begin{equation}
D_c =D_p\left(\frac{\rho_p U^2}{Y}\right).
\label{eq:strength_Dc}
\end{equation}
Then, the strength $Y=2.5$ MPa is obtained from the condition $D_p=D_c/2$. We can even estimate the finite strength of the target in this framework. The existence of the finite strength implies that the target material should somehow be viscoplastic. Substituting Eq.~(\ref{eq:strength_Dc}) to Eq.~(\ref{eq:U_RT}), we obtain a form of the impact velocity in strength-dominant cratering regime as, 
\begin{equation}
U =\left( \frac{12 N^2 \gamma}{\rho_f D_p}\right)^{3/10}\left(\frac{Y}{\rho_p}\right)^{1/5}.
\label{eq:strength_v0}
\end{equation}
Since this estimate is based on the simple scaling analysis, only the rough order estimate is possible. In terms of the order estimate, the above estimate might be reasonable. However, there are some severe problems that must be overcome to fully understand the impact causing the wavy nano craters rim. Such problems are briefly discussed below. 

In general, the region of fingering instability depends on surrounding gas pressure~\cite{Xu} and wettability of the projectile~\cite{Duez} as well as impact inertia and properties of projectile and target. The droplet impact under vacuum condition cannot cause the fingering~\cite{Xu}. If the surface of projectile is hydrophilic, no-splashing is observed until the critical impact inertia~\cite{Duez}. These characteristics relate to the transition from the no-fingering to the clear-fingering states. While these effects might influence the value of parameters at the transition point, we consider the scaling exponent would not be affected. Furthermore, a recent work has revealed that the fingering of a droplet impact depends on the shape of the target~\cite{Paulo}. In the current investigation, we performed a very simple experiment and found that $N$ might be scaled by the combination of $We$, $Fr$ and $\rho_r$. This means that the crater deformation and the associated Rayleigh-Taylor instability are the possible sources of the fingering pattern formation. 

While only the viscous fluid targets are used in this study, viscoelastic or plastic behavior of the target must be also studied in order to discuss the general fingering instability. Moreover, the finite strength is indeed necessary to explain the observed result as mentioned above. Elastic property may work as a resistance to the splashing just like an atmospheric pressure which induces the fingering instability for the droplet impact~\cite{Xu}. Such solid-like elastic or plastic property is advantageous also to retain the crater shape on the target surface just like Itokawa's nano crater. However, much higher impact velocity and vacuum conditions are needed to study a viscoelastic or viscoplastic fingering. This is the important future problem. 

\section{Summary}
We performed the low-velocity impact experiment with steel sphere projectiles and viscous target liquids. Milk-crown-like fingering can be observed in a particular parameter regime. The number of fingers created by the impact was counted from the high-speed video images. The observed fingers number $N$ can be scaled by the relation $N \simeq (\rho_r Fr)^{1/4}We^{1/2}/3^{3/4}$. This scaling is based on the Rayleigh-Taylor instability and the balance between the projectile's kinetic energy and the target cavity's potential energy. While the scaling reveals a certain aspect of the fingering induced by the impact to soft targets, further studies are necessary to clarify the details of impact fingering instability. Using a Rayleigh-Taylor-based scaling ($N \simeq D_c We^{1/2}/2\sqrt{3}D_p$), one can estimate the impact velocity for nano crater's wavy rim found on samples returned from the asteroid Itokawa. Assuming the strength-dominant cratering, the strength of the target material can also be estimated. However, more careful estimate by considering extreme conditions such as vacuum and low temperature is necessary to discuss the origin of wavy rim structures on Itokawa's nano craters.

\section*{Acknowledgments}
We would like to acknowledge S. Watanabe for introducing the Itokawa's nano crater shapes.
This research has been partly supported by the JSPS, KAKENHI, number 23654134 and number 26610113.

\end{document}